\documentstyle[pre,aps]{revtex}
\input epsf
\def\Sv{{\bf S}}
\def\pv{{\bf p}}
\def\Ev{{\bf E}}

\def\ev{{\bf e}}
\def\gradv{\nabla}

\def\rv{{\bf r}}
\def\Rv{{\bf R}}

\def\nv{{\bf n}}
\def\Pv{{\bf P}}
\begin{document}
\draft
\twocolumn[\hsize\textwidth\columnwidth\hsize\csname@twocolumnfalse\endcsname
\title{Topological Defects and Interactions in Nematic Emulsions}
\author{T.C. Lubensky, David Pettey, and Nathan Currier}
\address{Department of Physics and Astronomy, University of Pennsylvania,
Philadelphia, PA 19104, USA}
\author{Holger Stark}
\address{Institut f\"{u}r Theoretische und Angewandte Physik,
Universit\"{a}t Stuttgart, D-70550 Stuttgart, Germany}
\maketitle
\begin{abstract}
Inverse nematic emulsions in which surfactant-coated water droplets are
dispersed in a nematic host fluid have distinctive properties
that set them apart from dispersions of two isotropic fluids or
of nematic droplets in an
isotropic fluid.  We present a comprehensive theoretical
study of the distortions produced in the nematic host by the dispersed
droplets and of solvent mediated dipolar interactions between droplets that
lead to their experimentally observed chaining.  A single droplet
in a nematic host acts like a macroscopic hedgehog defect.  Global boundary
conditions force the nucleation of compensating topological defects in the
nematic host.  Using variational techniques, we show that in the lowest
energy configuration, a single water droplet draws a single hedgehog out of
the nematic host to form a tightly bound dipole.  Configurations in which
the water droplet is encircled by a disclination ring have higher energy.
The droplet-dipole induces distortions in the nematic host that lead to an
effective dipole-dipole interaction between droplets and hence to chaining.
\end{abstract}
\pacs{PACS numbers: 77.84.Nh, 61.30.Cz, 61.30.Jf}
\vskip2pc]
\narrowtext
\section{Introduction}
\par
Topological defects\cite{Kleman83,Mermin79,Trebin82,ChaLub95},
which are a necessary consequence of broken
continuous symmetry, exist in systems as disparate as superfluid
helium 3\cite{VolWol90} and 4\cite{WilBet87},
crystalline solids\cite{Taylor34,Friedel64,Nabarro67},
liquid crystals\cite{KurLav88,ChaRan86},
and quantum-Hall fluids\cite{SarPin97}.  They play an important
if not determining role in such phenomena as response to external
stresses\cite{Friedel64,Nabarro67},
the nature of phase transitions\cite{ChaLub95,Nelson83,Strandburg88},
or the approach to equilibrium after a quench into an ordered
phase\cite{Bray94}; and they are the primary ingredient in such phases
of matter as the Abrikosov flux-lattice phase of
superconductors\cite{Abrikosov57,BlaFei94} or the  twist-grain-boundary phase
of liquid crystals\cite{RenLub88-1,GooWau88-1,GooWau89-1}.
They even arise in certain cosmological models\cite{ChuDur91}.
Given the universal nature of topological defects, it is
always interesting to find new systems that allow us to increase our
understanding of these defects.  In this paper, we will present a detailed
theoretical investigation of a new class of nematic emulsions\cite{PouSta97}
whose intriguing properties are controlled by a class of topological defects
called hedgehogs.  These emulsions
are either simple inverse emulsions in which surfactant-coated
water droplets are dispersed in an aligned nematic host, or they are
multiple emulsions in which water droplets are dispersed in larger nematic
drops that in turn are dispersed in water.
\par
Liquid crystals are ideal materials for studying topological defects.
Distortions yielding defects are easily produced through
control of boundary conditions, surface geometries, and external fields.
The resulting defects are easily imaged optically.  The many different
liquid crystalline phases (nematic, cholesteric, smectic-$A$, smectic-$C$,
etc.) with different symmetry ground states make it possible to study
different kinds of defects.  Over the years, liquid crystals have provided
us with detailed and visually striking information about topological defects.
\par
Liquid crystal emulsions in which surfactant-coated drops containing a
liquid-crystalline material are dispersed in water have been a particularly
fruitful medium for studying topological
defects\cite{Meyer69,DuBPar69,CanLeR73,Drzaic95,KurLav88}.  The
liquid-crystalline drops are
typically from $10\mu$m to $50\mu$m in diameter and are visible
under a microscope.  Changes in alignment direction, specified
by the Frank director $\nv$, are easily studied under crossed polarizers.  The
isolated drops in these emulsions provide an idealized spherical confining
geometry for the liquid crystal.  More general distorted or multiply connected
random geometries\cite{Drzaic95} such as those produced in
polymer-dispersed liquid crystals
(PDLCs)\cite{DoaVaz86,CraZum96}, in emulsion films, or in dispersions of
agglomerations of silica spheres in a nematic host\cite{KreTsc92}
are of considerable current interest
because of display technologies based upon changing the light scattering
properties of these systems through modification of defect distributions via
external fields.
\par
In this paper, we will study inverse and multiple nematic emulsions.  These
emulsions differ from the direct emulsions described above in that isotropic
water droplets are dispersed in a nematic host rather than the other way
around. They are considerably more complex than direct emulsions.  In direct
emulsions, the nematic is separated into distinct, nearly spherical drops.
Normal or homeotropic boundary conditions on the nematic director at a drop's
surface will lead to a single point hedgehog defect in its
interior; tangential boundary conditions will lead to a pair of surface
defects called boojums\cite{Mermin77,CanLeR73,KurLav82}.
Though there can be transitions among
various director configurations as temperature or boundary conditions are
changed\cite{VolLav83,LavTer86}, the topological structure of these drops is
simple. In inverse emulsions,
each water drop with homeotropic boundary conditions will create a hedgehog
director configuration in its immediate vicinity.  Global boundary conditions
at the surface of the nematic restrict total topological charge.  Thus, in
order to satisfy global constraints, additional defects must be created out of
the nematic to compensate for or to cancel the topological charge created
by droplets. The nature and placement of these additional defects determine the
far-field director distortion produced by a droplet and the nature of
droplet-droplet interactions.  Experiments\cite{PouSta97,PouWei97} show that
each water droplet creates a companion point defect leading to dipole
distortions of the director field at large distances.  This is in contrast to
the quadrupolar ``Saturn-ring" configuration in which a disclination ring
encircles a droplet at its equator that has been extensively
studied\cite{Terentjev95,KukRuh96,RamRaj96,RuhTer97}.
Our calculations show that the experimentally observed dipole configuration
is the preferred one and that it leads to a dipole-dipole interaction between
drops that gives rise to the experimentally observed chaining of droplets.
It is interesting to note that similar topological dipole configurations
appear in two-dimensional systems including (1) free standing smectic
films\cite{LinCla97} where a circular region with an extra layer
plays the role of the
emulsion water droplet and (2) Langmuir films\cite{KnoRud97} in which a
liquid-expanded inclusion in a tilted liquid-condensed region plays a similar
role.
\par
The outline of this paper is as
follows.  In Sec. II, we review important elastic and topological properties of
nematics.  In Sec. III, we provide an overview of important experimentally
observed properties of inverse and multiple nematic emulsions.  In Sec. IV, we
calculate the director configurations and energy of a single water droplet in a
uniform nematic using various variational ansatzes.  In Sec. V, we introduce a
phenomenological free energy to describe long-distance director distortions
and interactions among droplets.  Finally, In Sec. VI, we summarize our
results.
\section{Order, Energy, and Topological Defects in nematics}
\par
A nematic liquid crystal is a uniaxial, homogeneous fluid characterized by a
unit vector $\nv$, called the Frank director, specifying the direction of the
principal axis of a symmetric-traceless-tensor order parameter.  The
ground-state free energy of a nematic is invariant under all spatially
uniform rotations of $\nv$ and under all inversions $\nv \rightarrow - \nv$.
In addition, all physically observable quantities are invariant under $\nv
\rightarrow - \nv$.  The ground-state manifold or order-parameter space is the
unit sphere in three dimensions $S^2$ with opposite points identified, i.e.,
the projective plane $RP^2 = S^2/Z_2$\cite{Mermin79,Trebin82,KurLav88}.
The topological structure of the
ground-state manifold determines the types of possible topological defects.
As we will review below, nematics can have both line defects (disclinations)
and point defects (hedgehogs).
\subsection{The Frank Free Energy}
The energy of slowly varying spatial distortions of the director $\nv ( \rv
)$ is determined by the Frank free energy
\begin{eqnarray}
F & = & \case{1}{2}\int d^3 r \{ K_1 ( \gradv \cdot \nv )^2 + K_2 (\nv \cdot
\gradv \times \nv )^2 \nonumber \\
& & \qquad+ K_3 [\nv \times (\gradv \times \nv ) ]^2\} \label{frank1}\\
& & -\int d^3r K_{24}\gradv \cdot [\nv \times (\gradv \times \nv
) + \nv ( \gradv \cdot \nv ) ] ,\nonumber
\end{eqnarray}
where $K_1$, $K_2$, $K_3$, and $K_{24}$ are, respectively, the splay, twist,
bend, and saddle-splay elastic constants.  (There is also the possibility
of another surface term with energy $K_{13} \gradv \cdot (\nv \gradv\cdot
\nv)$\cite{Oseen33,NehSau71}, which we will not consider in this
paper.) The saddle splay
term is a pure divergence; it reduces to integrals over all surfaces,
including interior surfaces formed, for example, by water droplets. For
spherical surfaces with normal boundary conditions, these integrals are
constant and do not vary, for example, when separations between droplets are
changed. To keep our calculations as simple as possible, we will use the
one-constant limit of the Frank free energy:
\begin{eqnarray}
F& = & \case{1}{2} K \int d^3 r [(\gradv\cdot \nv)^2 + (\gradv \times \nv )^2
] \label{frank2} \\
& & - K_{24} \int d\Sv \cdot [\nv \gradv \cdot \nv + \nv
\times (\gradv\times\nv )] \nonumber \\
&= & \case{1}{2} K \int d^3 r \nabla_i n_j \nabla_i n_j \label{frank2a}\\
& &  +\case{1}{2} (K -
2K_{24} ) \int d \Sv \cdot[\nv \gradv \cdot \nv +
\nv \times(\gradv\times\nv )] . \nonumber
\end{eqnarray}
Since surface energies do not play an important role in the phenomena to be
discussed in the paper, we will set the saddle-splay constant $K_{24}$ equal
to zero unless otherwise specified. When $K=2 K_{24}$, the free
energy reduces to the first line of Eq.\ (\ref{frank2a}), which is invariant
with respect to rigid rotations of any director configuration.  [Note:  In Ref.
\onlinecite{PouSta97}, calculations were done with $K=K_{24}$.]
\subsection{Surface Energies}
\par
Surfaces generally impose a preferred alignment direction of the nematic
director relative to their local normals.  The energetics of this alignment
are described by the Rapini-Papoular phenomenological surface free
energy\cite{RapPap69}
\begin{equation}
F_S =  \case{1}{2}W \int dS \sin^2 \gamma
\end{equation}
where $\gamma$ is the angle between the director and the surface normal.
Homeotropic or normal alignment is favored by $W>0$ and tangential alignment
by $W<0$.  The coupling constant $W$ varies in the range $10^{-4}$-$1$
erg/cm$^2$\cite{BliKab89} with typical values of order $10^{-3}$
erg/cm$^2$\cite{ErdZum90}.
\par
In addition to the above surface alignment energy, in emulsions, there is the
energy arising from the surface tension of the water-surfactant-oil
interface.  This energy is simply the surface tension $\sigma$ times the
total surface area
\begin{equation}
F_{\sigma} = \sigma \int dS .
\end{equation}
The surface tension is of order $10$erg/cm$^2$\cite{MasKro96}.
\par
We can now discuss the relative importance of the surface energies and the
bulk Frank energy.  Consider a spherical nematic drop of radius $a$ with
$W>0$.  If the director is everywhere normal to the surface, as the surface
alignment energy favors, the Frank elastic energy is $8 \pi K a$, and the
surface alignment energy is zero. On the other hand, if the
director is parallel throughout the interior of the drop, the Frank energy is
zero, and the surface alignment energy is $8 \pi W a^2/3$.  The surface
energy scales as $a^2$, whereas the elastic energy scales as $a$.  Thus,
surface energy dominates over elastic energy for large drops, and we may
assume, to a good approximation, that the preferred direction of surface
alignment is imposed as a constraint.  On the other hand for small droplets,
elastic energy dominates over surface energy, and we should expect the
surface director to deviate from its preferred orientation. The
characteristic droplet dimension beyond which we may assume rigid boundary
conditions is $r_c = K/W \approx 10^{-6}/3 \times 10^{-2}\approx 0.3\mu$m.
Typical droplet radii in the experiments of Poulin {\it et al.}\cite{PouSta97}
are larger than $1\mu$m, and we may use rigid boundary conditions to
interpret them.
\par
Similar considerations apply to shape distortions of the droplets.  The
positive surface tension favors spherical drops of either liquid crystal in
water or of water in liquid crystal. The surface energy scales as $\sigma
a^2$.  Thus, we can expect drops to be spherical and undistorted by
the nematic director for drops larger than $r_{\sigma} = K/\sigma \approx
1$nm. In what follows, we will assume that water droplets remain spherical
and that normal boundary conditions are rigidly imposed at nematic-water
interfaces.
\subsection{Topological Defects\label{Topdefects}}
\par
\begin{figure}
\centerline{\epsfbox{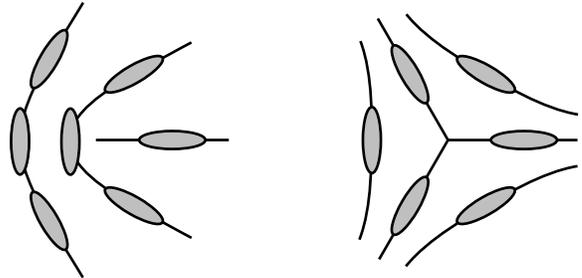}}
\caption{Two director configurations for a strength $1/2$ disclination.
In a two-dimensional nematic, the left and right figures correspond,
respectively, to disclinations of strength $+1/2$ and $-1/2$. In a
three-dimensional nematic, these
configurations can be converted into each other via continuous
transformations of the director}
\label{figdisc1}
\end{figure}

Topological defects in ordered media are singular regions of spatial
dimension less than that of physical space that are surrounded by
order-parameter configurations that cannot be transformed to a homogeneous
ground state via continuous deformations.  There are two kinds of topological
defects in a nematic.  They are (1) line defects, called disclinations, with
winding number of strength $1/2$ in which the director undergoes a rotation
of $\pi$ in one circuit around any one-dimensional path encircling the linear
defect core, and (2) point defects, called hedgehogs, in which the director
sweeps out all directions on the unit sphere $S^2$ as all points on any
two-dimensional surface enclosing the defect core are visited.  The only
topologically stable disclinations have winding number
$1/2$.  All director configurations on a loop can either be shrunk
continuously to a single point in the order parameter space, in which case
the loop encloses no defect, or they can be continuously distorted to a path
in $RP^2$ starting at some arbitrary point and ending at a diametrically
opposite point, in which case the loop, encloses a disclination of strength
$1/2$.  Typical director configurations for a strength $1/2$ disclination are
shown in Fig. \ref{figdisc1}.
\par
In the simplest disclination configurations shown in Fig. \ref{figdisc1}, the
director is $\nv = (\cos \phi/2 , \pm\sin \phi/2 , 0)$, where $\phi = \tan^{-1}
y/x$ is the azimuthal angle in the $xy$-plane.  The energy per unit length of
such disclination lines calculated from Eq. (\ref{frank2}) is
\begin{equation}
\epsilon = \case{1}{4} \pi K \ln ( R/r_c ) + \epsilon_c ,
\label{discenergy}
\end{equation}
where $R$ is the sample radius, $r_c$ is the radius of the
disclination core, and $\epsilon_c$ is the core energy per unit length, which
is of order $K$.
\par
Hedgehogs are point defects characterized by an integer topological charge $q$
specifying the number of times the unit sphere is wrapped by the director on
any surface enclosing the defect core. An analytical expression for $q$
is\cite{Trebin82}
\begin{equation}
q = {1 \over 8 \pi} \int dS_i \epsilon_{ijk} \nv \cdot (\partial_j \nv \times
\partial_k \nv ) ,
\label{topcharge}
\end{equation}
where the integral is over any surface enclosing the defect core.
For an order parameter with $O_3$ (vector) symmetry, the order-parameter
space is $S^2$, and hedgehogs can have positive or negative charges.  Nematic
inversion symmetry makes positive and negative charges equivalent, and we may,
as a result, take all charges to be positive.
\par
\begin{figure}
\centerline{\epsfbox{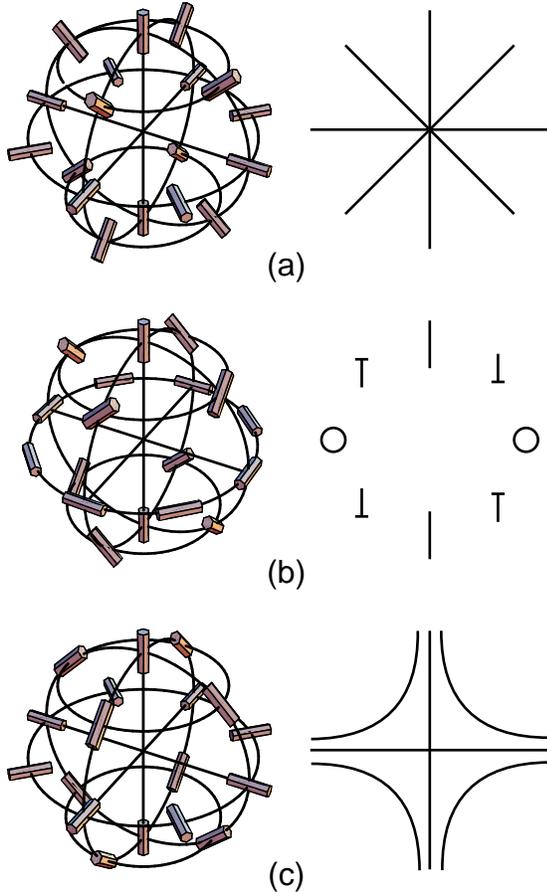}}
\caption{(a) A radial hedgehog in which the director points radially
outward from a central point like the electric field of a point charge.
(b) A circular hedgehog obtained from a radial hedgehog by rotating the
director at every point
through $\pi/2$ about the vertical axis. (c) A hyperbolic hedgehog
obtained from the radial hedgehog by rotating the director at every point
by $\pi$ about the vertical axis. In each case, the figures at the left
provide a three-dimensional depiction of the defect whereas that at the right
shows a projection onto any plane containing the polar axis.  In (b),
standard notation in which the nail heads indicate the end of the director
coming out of the plane is used.}
\label{fighedge1}
\end{figure}

There is a continuous infinity of director configurations for each value of
the hedgehog charge.  In the simplest unit-charge hedgehog configuration
shown in Fig. \ref{fighedge1}(a),
the director points radially outward from the point core like the electric
field near a point charge. This configuration is called a {\it radial}
hedgehog for obvious reasons.  Other configurations can be obtained from the
radial configuration via rotations through arbitrary angles about any axis.
Two examples are show in Figs. \ref{fighedge1}(b) and (c). When the
director of a radial hedgehog is rotated about a fixed axis
through $\pi$, a {\it
hyperbolic} hedgehog shown in Fig. \ref{fighedge1}(c) is produced.  The
hyperbolic hedgehog can be obtained from a radial hedgehog via a series of
continuous distortions of the director passing through intermediate
configurations such as the ``circular" configuration shown in Fig.
\ref{fighedge1} (b).  Thus, radial, hyperbolic, and all intermediate hedgehogs
are topologically equivalent.
\par
The energies of the simple hedgehog configurations shown in Fig.
\ref{fighedge1} in a sphere of radius $R$ with free boundary conditions at
the outer surface are easily calculated from the Frank free energy [Eq.\
(\ref{frank1})]. The Frank director for these configurations are
$\nv = (x,y,z)/r$ for the radial, $\nv = (y,-x,z)/r$ for the circular, and
$\nv = (-x,-y,z)/r$ for the hyperbolic hedgehogs, where $\rv = (x,y,z)$ and $r
= |\rv |$. In a spherical region of radius
$R$, their respective energies are
\begin{eqnarray}
E_{\rm radial} & = & 8 \pi (K_1 - K_{24}) R \nonumber \\
&\rightarrow &
8 \pi (K - K_{24}) R\nonumber \\
E_{\rm circ}&=&{8 \pi\over 15}(3 K_3 + 5K_2 + 2 K_1 - 5K_{24})R\nonumber \\
&\rightarrow& {8 \pi \over 3}(2 K - K_{24})R \nonumber \\
E_{\rm hyper}& = & {8 \pi\over 15}(3 K_1 + 2K_3 + 5K_{24} ) R
\nonumber\\ &\rightarrow &{8\pi \over 3 }( K + K_{24}) R ,
\label{hedgeenergy}
\end{eqnarray}
where the final expressions are for the case of equal elastic constants.
When $K_{24} = 0$, these energies reduce to those calculated in Ref.
\onlinecite{LavTer86}. The hyperbolic hedgehog has lower energy than the radial
hedgehog provided $K_3 < 6 K_1 - 10 K_{24}$ or $K> 2 K_{24}$ for single
elastic constant approximation.  Thus, if $K_{24}
= 0$, the hyperbolic hedgehog always has the lower energy.  The circular
hedgehog has the most bend.  Since $K_3$ is generally the largest elastic
constant, the circular hedgehog generally has the highest energy provided
$K_{24}$ is not too large. If $K= 2K_{24}$, the energies of the three hedgehog
configurations are equal (and equal to $4 \pi K R$), as one could have
predicted from Eq.\ (\ref{frank2}), which is invariant with respect to rigid
rotations of even a spatially varying
$\nv$ when $K= 2 K_{24}$.  In confined geometries, the Rapini-Papoular surface
energy competes with the
$K_{24}$ surface term to determine defect configurations.
\par
In systems with vector symmetry,
the combined topological charge [i.e., the charge obtained by evaluating Eq.
(\ref{topcharge}) on any surface enclosing both hedgehog cores] of two
hedgehogs with respective charges $q_1$ and $q_2$ is simply the sum $q_1 +
q_2$.  In nematics, the sign of the
topological charge has no meaning, and the combined topological charge of
two hedgehogs is either $|q_1 + q_2 |$ or $| q_1 - q_2 |$.  It is impossible
to tell with certainty which of these possible charges is the correct
one by looking only at surfaces enclosing the individual hedgehogs.
\par
\begin{figure}
\centerline{\epsfbox{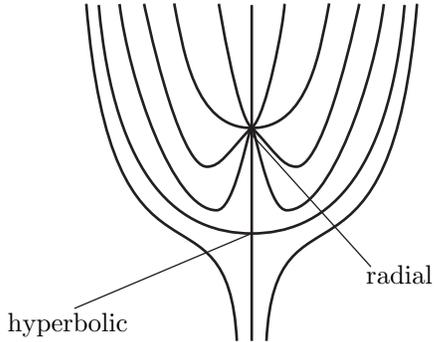}}
\caption{A radial and a hyperbolic hedgehog combining to give a
configuration with hedgehog charge zero}
\label{fighedge2}
\end{figure}

We will be primarily interested in how two unit-charge hedgehogs can combine
to give a hedgehog charge of zero.  Figure \ref{fighedge2} shows how a radial
and a hyperbolic hedgehog can combine to give a charge-zero configuration,
i.e., a configuration in which the director is parallel at infinity.
\par
\begin{figure}
\centerline{\epsfbox{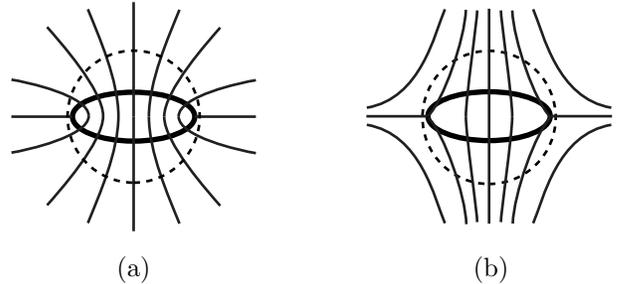}}
\caption{Disclination rings with unit hedgehog charge: (a) radial hedgehog
and (b) hyperbolic hedgehog. The dotted line in each figure represents a
sphere of radius $R$ beyond which the director configuration is that of a
hedgehog.}
\label{figdiscring}
\end{figure}

Disclination rings can carry a hedgehog charge $q$ as measured by the
integral in Eq.\ (\ref{topcharge}) evaluated over a surface enclosing the
ring\cite{Garel78,NakHay88}. Figure \ref{figdiscring}
depicts disclination rings with
far-field director configurations corresponding to radial and hyperbolic
charge $1$ hedgehogs. These rings can be shrunk to a point leaving a point
hedgehog.  Since the disclination ring is topologically equivalent to a
hedgehog, one can ask whether it is energetically favorable for a point
hedgehog to open up to a disclination
ring\cite{MorNak88,Terentjev95}. If one
assumes that order parameter configurations remain uniaxial,
one can obtain a crude
estimate of the radius $R_0$ of the disclination ring using the expressions,
Eq. (\ref{discenergy}) and (\ref{hedgeenergy}), for disclination and hedgehog
energies.  The director configuration of a charge $1$ disclination ring is
essentially that of simple disclination line discussed above Eq.
(\ref{discenergy}) in the vicinity of the disclination core, i.e., up to
distances of order $R_0$ from the ring center.  Beyond this radius, the
director configuration is approximately that of a hedgehog (radial or
hyperbolic).  Thus, we can estimate the energy of a disclination ring of
radius $R_0$ centered in a spherical region of radius $R$ to be
\begin{equation}
E_{\rm ring} \approx 2 \pi R_0 [\case{1}{4} \pi K \ln (R_0/ r_c) +
\epsilon_c] + 8\pi \alpha K (R - R_0 ) ,
\end{equation}
where $\alpha = 1-k_{24}$ for a radial hedgehog and $\alpha = (1 + k_{24})/3 $
for a hyperbolic hedgehog, where $k_{24} = K_{24}/K$.  Minimizing over $R_0$
and
setting
$\epsilon_c = K$, we find
\begin{equation}
R_0 = r_c \exp\left[{16 \over \pi}
\left( \alpha - {1 \over 4} - {\pi\over 16}\right)\right] .
\label{R0}
\end{equation}
Though admittedly crude, this approximation gives a result that has the same
form as that calculated in Refs. \onlinecite{MorNak88,Terentjev95,LavTer97}
using a more sophisticated continuous ansatz.
It has the virtue that it applies to
both radial and hyperbolic far-field configurations. It predicts that the
hedgehog with the lower energy far-field configuration (i.e., the one
with smaller $\alpha$) will have the smaller disclination-ring radius.  If
$k_{24} = 0$, the hyperbolic hedgehog has the lower energy with $\alpha =
1/3$ rather than $\alpha =1 $.  In this case, the core of a radial hedgehog
should be a ring with radius $R_0 \approx r_c e^{2.8}$, or $R_0 \approx 0.2
\mu$m for $r_c \approx 100 \AA$.  The core of the hyperbolic hedgehog, on the
other hand will be a point rather than a ring because $R_0 \approx r_c
e^{-0.58} < r_c$.
\par
If the constraint that the tensor nematic order parameter $Q_{ij}$ be
uniaxial is relaxed, then the core of a disclination can become
biaxial\cite{SchSlu87} with a core radius of order the biaxial correlation
length $\xi_b$.  The energy of a disclination is still given by Eq.\
(\ref{R0}) with $r_c \sim \xi_b$ and with a core energy determined
by the energy difference between the biaxial and uniaxial state rather than
the energy difference between the isotropic and nematic states.  A hedgehog
can also develop a biaxial core with radius of order $\xi_b$.  Because the
biaxial core is characterized by a non-vanishing biaxial order parameter,
its structure is not the same as that of the uniaxial disclination ring
discussed above.  Calculations\cite{PenTre89} based on the
Landau-de Gennes free energy for a nematic predict a biaxial core size of
order $0.025\mu$m for MBBA.  A detailed analysis of the competition between
a biaxial core and a biaxial disclination ring has not been done.
\par
It is very difficult to predict with certainty what the
core structure of a hedgehog will be. If the core is a disclination ring, its
radius varies exponentially with the elastic constants.  If the core is
biaxial, it will have a biaxial structure out to a radius of order the
biaxial correlation length, which should be of order $100$\AA\ or less.  The
general arguments given above would lead one to expect hyperbolic hedgehogs
to have the smallest core size.  In the experiments of Poulin
{\it et al.}\cite{PouSta97,PouWei97}, all hyperbolic hedgehogs that were
observed have cores that are point-like to the resolution of the optical
microscope.
\section{Director Configurations in Inverted Nematic Emulsions}
\par
\begin{figure}
\centerline{\epsfbox{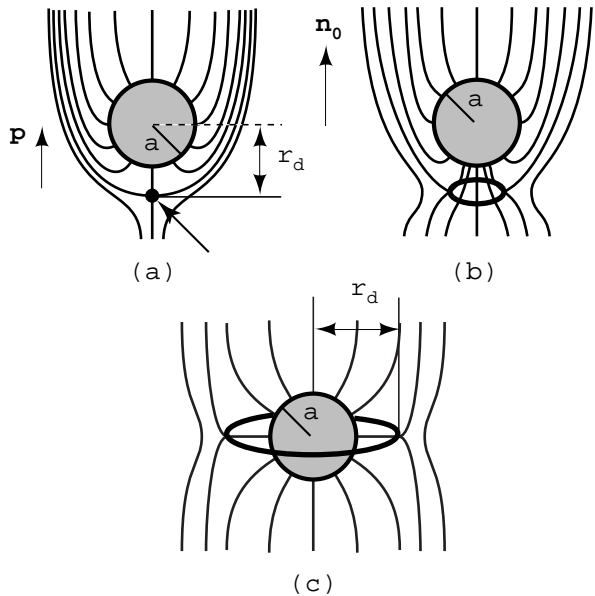}}
\caption{Possible director
configurations induced by a single spherical
droplet with homeotropic boundary conditions in a nematic with total
topological charge of zero. (a) Dipole configuration with a companion
hyperbolic hedgehog (indicated by an arrow).
(b) Dipole configuration with a companion hyperbolic
disclination ring. (c) Quadrupolar saturn ring configuration with a
disclination ring encircling the water droplet at its equator. The direction
of the topological dipole $\pv$ is shown in (a).}
\label{figdroplet1}
\end{figure}
\begin{figure}
\centerline{\epsfbox{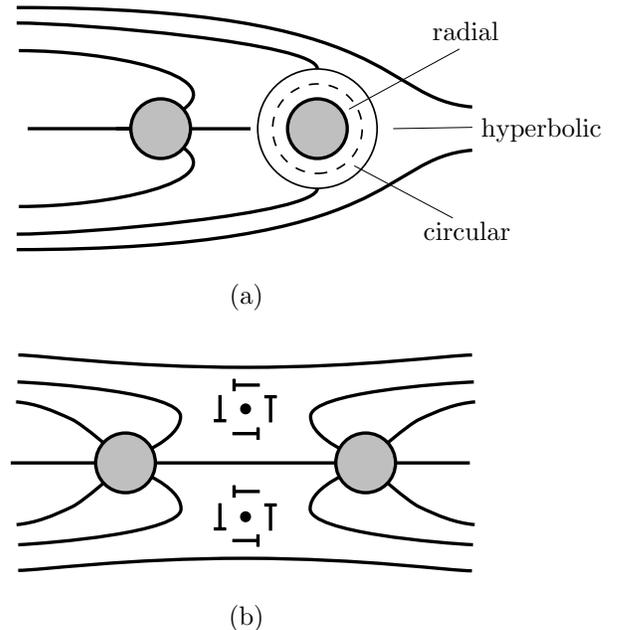}}
\caption{Schematic representation of the nonsingular
director configuration produced by two water droplets whose boundary
conditions produce radial hedgehogs.
(a) The radial hedgehog around one droplet converts
continuously to a hyperbolic configuration which then combines with the
radial configuration of the other droplet.  (b) The radial configuration
around each droplet converts smoothly to a toroidal ``escaped strength one"
non-topological disclination that encircles the axis defined by the
droplets. We are grateful to R.B. Meyer for suggesting configuration (b) to
us.}
\label{figdroplet2}
\end{figure}

In the experiments reported by Poulin, {\it et al.}\cite{PouSta97}, a nematic
liquid crystal (pentyl cyano biphenyl, or 5CB), a surfactant (sodium dodecyl
sulfate, SDS) and water are mixed together to produce inverted and multiple
liquid crystalline emulsions.  The inverted emulsions are placed in a thin
rectangular cell of approximate dimensions
$20\mu{\rm m} \times 1 {\rm cm} \times
1 {\rm cm}$. The large-area upper and
lower surfaces were treated to produce tangential boundary conditions.  Thus,
the total hedgehog charge $Q$ in the cell, obtained by performing the
integral in Eq.\ (\ref{topcharge}) is zero.  With normal boundary conditions,
each water droplet nucleates a radial hedgehog of charge one.  To maintain
zero charge in the cell, compensating director distortions, usually point
or line defects, must be created
out of the nematic itself.  Possible director configurations of a single
droplet with total charge zero are shown in Fig. \ref{figdroplet1}. A single
droplet could nucleate a companion hyperbolic hedgehog (Fig.\
\ref{figdroplet1}a), or it could nucleate a disclination ring of finite
radius lying above or below the droplet (Fig.\ \ref{figdroplet1}b) or
encircling the droplet in a ``Saturn-ring" configuration (Fig.\
\ref{figdroplet1}c).  Director configurations for many droplets can be
constructed from the single director configurations shown in Fig.\
\ref{fighedge1}.
Other configurations in which the hedgehog charge of water droplets is
canceled by continuous textures in the surrounding nematic rather than by
the formation of point hedgehogs or singular disclination rings are
possible.  For example, if there are two droplets, the radial
configuration around one droplet could continuously deform to a hyperbolic
configuration passing through intermediate configurations such as the
``circular" hedgehog of in Fig.\ \ref{fighedge1}b.
The final hyperbolic configuration could
combine with the radial configuration of the neighboring droplet to produce
a configuration with zero charge but without any point defects in the
nematic as shown in Fig. \ref{figdroplet2}a.
Alternatively, there could be a more symmetric configuration with a
toroidal ``escaped strength one" non-topological
disclination line\cite{Meyer73} as shown in Fig.\ \ref{figdroplet2}b.
\par
\begin{figure}
\centerline{\epsfbox{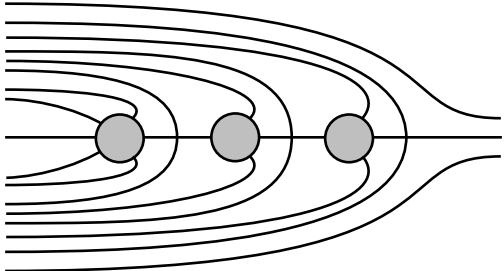}}
\caption{Schematic representation of a chain of three water droplets in a
cell with parallel boundary conditions at infinity.  Each droplet creates a
companion hyperbolic hedgehog, and droplets and companions defects lie on a
single line.}
\label{figexpt1}
\end{figure}

In the experiments of Poulin {\it et al.}\cite{PouSta97,PouWei97}, the
dipole configuration shown in Fig.\ \ref{figdroplet1}a is almost always
observed.  When
many droplets are in the cell, each droplet forms a dipole with a companion
hyperbolic defect so the total charge of the multiple droplet system is
zero as required.  Furthermore, the droplet-dipoles align in chains
parallel to the cell-director as shown schematically in Fig.\ \ref{figexpt1}
(See also Figs. 8 and 9 of Ref. \onlinecite{PouWei97}).
Occasionally, droplet pairs are observed to induce director configurations
that cannot be interpreted in terms of companion hedgehog defects [See Fig.
12 of Ref. \onlinecite{PouWei97}].  These configurations may be of the type
shown in Fig.\ \ref{figdroplet2}b.
\par
\begin{figure}
\centerline{\epsfbox{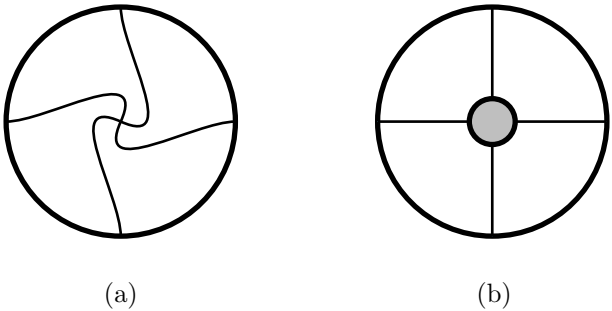}}
\caption{(a) Schematic representation of the director configuration of a
nematic drop with no interior water droplet.  the director is forced by
boundary conditions to have a radial configuration at the out surface.  As
distance from the surface increases, the director seeks lower energy,
nonradial configurations. (b) Schematic representation of the director
configuration of a nematic drop with a single interior water droplet.
Homeotropic boundary conditions at the outer and water-droplet surfaces
force a radial configuration everywhere.}
\label{figexpt2}
\end{figure}
\begin{figure}
\centerline{\epsfbox{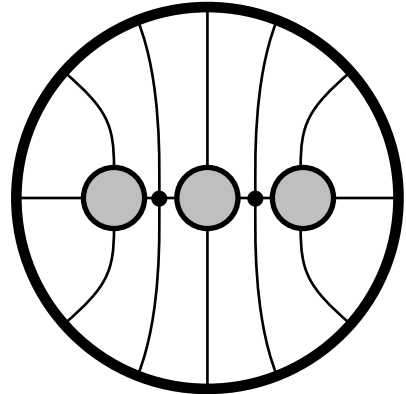}}
\caption{A nematic drop with 3 internal water droplets.  The 3 water
droplets and their 2 companion hyperbolic defects form a linear chain
at the center of the nematic drop. The total charge of this configuration
is one.}
\label{figexpt3}
\end{figure}

In multiple emulsions, water droplets are confined to the interior of nematic
drops with spherical symmetry.  If the outer surface of the nematic drop
enforces homeotropic boundary conditions, then the total topological charge
in the nematic drop is one.  If there are no water droplets in the nematic
drop, there must be a point hedgehog defect in the interior of the drop.
In general, the radial hedgehog favored by homeotropic boundary conditions
at the outer surface does not have the lowest energy.  Instead, there is an
evolution away from the radial configuration with distance from the droplet
surface\cite{PreArr74,LavTer86}, as depicted in Fig.\
\ref{figexpt2}a.  Under crossed polarizers, this configuration will appear
as rotating cross.
A single water droplet in the interior of the nematic drop will create
a radial hedgehog.  Since the total topological charge of the nematic drop is
one, no compensating defects must be created from the nematic.  The
configurations enforced at the water droplet surface and at the outer surface
of the nematic drop are both radial.  As a result, the director adopts a
radial configuration throughout the drop
as depicted in Fig.\ref{figexpt2}b (See Fig. 13 of Ref.
\onlinecite{PouWei97}).  Under crossed polarizers, this
configuration will appear as a rigid unrotated cross.
A second droplet added to a nematic drop
creates an additional interior radial hedgehog. In order to satisfy the global
boundary condition of charge one, a hyperbolic defect is created out of the
nematic.  If there are $N$ water droplets inside a nematic drop, $N-1$
hyperbolic defects will be created. The droplets and defects form linear
chains with an unpaired droplet as shown in Fig.\
\ref{figexpt3} (See also Fig. 14 of Ref. \onlinecite{PouWei97}).
These chains (or the single water droplet if that is all
there is) are rigidly placed at the center of the nematic droplet and
undergo no observable Brownian motion.
\section{Configuration and Energy of Single Droplet}
\par
In the preceding section, we discussed possible director configurations
induced by the presence of spherical water droplets with
homeotropic boundary conditions
in a nematic with parallel boundary conditions at
infinity. Experiments show that the water
droplets create companion hyperbolic hedgehogs rather than disclination rings.
In this section, our goal is to calculate the equilibrium separation of the
droplet from its companion and to compare the energy of the dipole
configuration with that of the saturn ring and intermediate configurations
depicted in Fig. \ref{figdroplet1}.  The calculational program is in principle
quite simple: solve the Euler-Lagrange equations for the director arising from
the minimization of the Frank free energy [Eq. (\ref{frank2})] subject to
the normal boundary conditions at the surface of the water droplet
and parallel boundary conditions at infinity.  Unfortunately, the
Euler-Lagrange equations are highly nonlinear, and analytical solutions cannot
be found except for a few special geometries and boundary conditions.
We can, however, obtain analytical solutions
for the director far from the droplet.  Using these solutions and
information, evident from Fig.\ \ref{figdroplet1}, about the form of
director configurations near the droplet, we can construct variational
ansatzes for the director that obey all boundary conditions and that have the
desired defect structure.  In this section, we will first discuss the nature
of the far-field solutions.  We will then use two different ansatzes to
calculate director configurations and their associated energies.  The first
ansatz applies only to the dipolar configuration.  The second applies to all
of the configurations in Fig.\ \ref{figdroplet1} and will allow us to compare,
for example, the energies of the dipolar and Saturn-ring configurations.
\subsection{Far-field Solutions}
\par
The constraint of zero topological charge requires $\nv ( \rv )$ to approach
$\nv_0 = ( 0,0,1)$ as $r \rightarrow \infty$. We assume that $\nv_0$ is
along the positive $z$ axis.  No physical results will change, however,
if we reflect $\nv_0$ to be along the negative $z$ axis.  At large but not
infinite
$r$, the deviation of $\nv ( \rv )$ from $\nv_0$ is small, and $\nv ( \rv )
\approx ( n_x , n_y , 1)$.  Thus, at large $r$, we can replace the full
nonlinear Frank free energy by the harmonic free energy
\begin{equation}
F_{\rm har} = \case{1}{2}K \sum_{\mu = x,y} \int d^3 r (\gradv n_{\mu} )^2 ,
\label{frankhar}
\end{equation}
where we introduced the notation $n_{\mu}$, $\mu = x,y$ for the components
of $\nv$ perpendicular to $\nv_0$. The Euler-Lagrange equations arising
from this equation are simply Laplace equations:
\begin{equation}
\nabla^2 n_{\mu} = 0 .
\end{equation}
At large $r$ the solutions to this equation can be expanded in multipoles:
\begin{equation}
n_{\mu} = {A^{\mu} \over r} + {{\bf p}^{\mu} \cdot \rv \over r^3} +
{c_{ij}^{\mu} r_i r_j \over r^5} + \cdots .
\label{multipole1}
\end{equation}
The solutions we seek are invariant
with respect to rotations about the $z$ axis and have no azimuthal
component to $\nv$ (i.e., no twist in $\nv$ about the $z$ axis).
This implies that
$A^{\mu} = 0$ and that $n_x$
and $n_y$ must be proportional, respectively, to $x$ and $y$.  In
addition, the dipolar part should change sign if the position of the
companion defect is shifted from above to below the droplet. The
requirements are met by setting $\pv^{\mu} = (\pv \cdot \nv_0 )
\ev^{\mu}$ and $c_{ij}^{\mu} =  c (n_{0i} e^{\mu}_j +
e^{\mu}_i n_{0j})$ where $e^{\mu}_i = \delta^{\mu}_i$ is the unit vector
pointing in the $\mu = x,y$ direction.  We identify the vector $\pv$ as the
dipole moment of the droplet-defect configuration, and $\pv\cdot \nv_0$
with its $z$ component. Thus if $\pv$ changes sign relative to
$\nv_0$, the dipole contribution to $n_{\mu}$ also changes sign.  In the
configurations we consider in this section, $\pv$ is aligned either
parallel or anti-parallel to $\nv_0$ so that $\pv \cdot \nv_0 = \pm p$
where $p$ is the magnitude of the dipole moment.  The parameter $c$, as
we will show in more detail in the next section, is the amplitude of
the quadrupole moment tensor $c_{ij}$ of the droplet-defect combination. Thus,
we have
\begin{eqnarray}
n_x &  = & p_z {x \over r^3} + 2c {z x \over r^5} \nonumber \\
n_y & = & p_z {y\over r^3} + 2 c {z y \over r^5} .
\label{multipole2}
\end{eqnarray}
By dimensional analysis, $p_z \sim a^2$ and $c\sim a^3$, where $a$ is the
radius of the sphere. Equations \ref{multipole2} produce the far-field
configurations of Fig.\ \ref{figdroplet1}a if we choose $p_z$ to be
positive when the companion hedgehog is below the droplet.
Thus, we adopt the convention that
the dipole moment of the droplet and its companion defect points from the
companion to the droplet.
\par
The multipole expansion of Eqs.\ (\ref{multipole1})
and (\ref{multipole2}) eventually breaks down because of nonlinearities
neglected in Eq.\ (\ref{frankhar}).  We can determine the leading corrections
by
including the leading anharmonic corrections to the harmonic free energy.  Far
from the defect, we can set $\nv = (n_x, n_y , \sqrt{1 - n_{\perp}^2}
) \approx (n_x, n_y , 1 - \case{1}{2} n_{\perp}^2 )$, where $n_{\perp}^2 =
n_{\mu} n_{\mu}$.  The leading anharmonic correction to $F_{\rm har}$ is
then
\begin{equation}
F_{\rm an} = \case{1}{8} K \int d^3r (\gradv n_{\perp}^2 )^2 ,
\end{equation}
and the Euler-Lagrange equations with this correction are
\begin{equation}
\nabla^2 n_{\mu} + \case{1}{2} n_{\mu} \nabla^2 n_{\perp}^2 = 0 .
\end{equation}
Using this equation, one can show that if the leading contribution to
$n_{\mu}$ is dipolar, then the first correction to $n_{\mu}$
arising from nonlinear terms is of the form $r_{\mu}/r^7$.  In other
words the multipole expansion of the Laplacian operator gives the
correct large $r$ behavior up to order $r^{-5}$.  Thus, we could in
principle develop variational approximations in which all of the
multipole moments from order $2$ to order $5$ are variational
parameters.  We will content ourselves with allowing only the dipole
and quadrupole moments to vary.
\subsection{The Electric-Field Ansatz}
\par
\begin{figure}
\centerline{\epsfbox{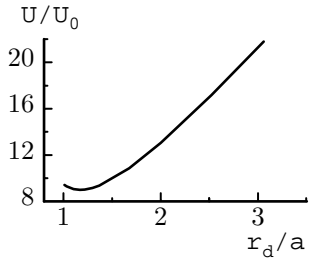}}
\caption{Energy (in units of $U_0= \pi K a/2$) of droplet dipole as a function
of the distance (in units of the droplet radius $a$) from the droplet center
to the companion hedgehog.}
\label{figEfield1}
\end{figure}

Any ansatz for $\nv$ for the dipole configuration of Fig.\
\ref{figdroplet1}a must be normal to the water droplets at $r=a$, tend
to $\nv_0$ as $r\rightarrow \infty$, and have a hyperbolic hedgehog at
some position along the $z$ axis outside of the water droplet.  The familiar
electrostatics problem of a charged conducting sphere in an external electric
field can provide the basis for an ansatz for $\nv$ that satisfies all of
these conditions.  The electric field $\Ev$ is normal to the conducting
sphere, and it tends to a constant $\Ev_0 = E_0 \ev_z$ as $r\rightarrow
\infty$.  If the charge $Q$ on the sphere is large enough, there is a point
below the sphere at which the electric field vanishes.  The normalized
electric field
configuration in the vicinity of this point is identical to that of a unit
vector in the vicinity of a hyperbolic hedgehog.  Thus, we have all of the
ingredients we need for a variational ansatz.  We have merely to choose
\begin{equation}
\nv ( \rv ) = \Ev( \rv ) /|\Ev ( \rv ) | .
\label{nelec}
\end{equation}
The electric field for the above problem is
\begin{equation}
{\Ev (\rv ) \over E_0} = \ev_z + \lambda^2 a^2 {\rv \over r^3} - {a^3 \over
r^5} ( r^2 \ev_z - 3 z \rv ) ,
\label{electric1}
\end{equation}
where $\lambda^2 = Q/(E_0 a^2)$ is a unitless measure of the strength of the
electric field produced by the charge $Q$ compared to the fixed external
field $E_0$ and $a$ is again the radius of the sphere.
The last term in this expression arises from an image dipole at the center
of the sphere that enforces the boundary condition that $\Ev$ be normal to
the surface of the sphere at $r=a$.
For $\lambda^2>3$, we find precisely one zero of the electric field
at $\rv = - z_0 \ev_z$ outside the sphere, where $z_0$ is the appropriate
solution to
\begin{equation}
|z|^3 - |z| \lambda^2 a^2  + 2 a^2 = 0 .
\end{equation}
(For $\lambda^2 = 3$, the point of zero electric field just touches the
sphere, and for $\lambda^2<3$, a singular ring appears on the surface of
the sphere.) $z_0$ is the distance $r_d$ from the droplet center to its
companion defect. At large $r$, $\nv ( \rv)$ becomes
\begin{equation}
n_{\mu} = (\lambda a )^2 {r_{\mu} \over r^3} + 3 a^3 {z r_{\mu} \over r^5}
\end{equation}
in agreement with Eq.\ (\ref{multipole2}).  Thus the dipole moment is
$\lambda^2 a^2$ and the quadrupole moment is $3 a^3/2$.  The variable
$\lambda$ is a variational parameter that determines both the position of
the hyperbolic defect and the magnitude of the dipole moment.
The ansatz fixes the quadrupole moment independent of the value of
$\lambda$ and constrains the dipole moment to be greater than $3a^2$.
A natural energy scale is $U_0 = \pi K a/2$.
The reduced energy $U/U_0$ calculated from Eqs. (\ref{nelec}),
(\ref{electric1}), and (\ref{frank1}) is plotted as a function of
the distance between the sphere and the companion defect in
Fig.\ \ref{figEfield1}. The energy at the minimum of the curve is
$U = 9.00 U_0$.  At this minimum, the other parameters characterizing the
droplet-defect pair are
$z_0 = 1.19 a$, $p_z = 3.02 a^2$, and $c = 3 a^2 /2$.
\par
Of course we have no a priori reason to expect a normalized electric
field to yield the true minimum energy configuration for this problem.
By modifying our chosen $\Ev(\rv)$ to be a slightly more general
vector field and by no longer insisting that it be a true electric field,
we may hope to improve our ansatz and to relax the constraints put upon
$p_z$ and $c$.  We introduce below one such generalization that provides a
significant improvement:  $U=
7.87 U_0$, $z_0=1.26a$, $p_z = 2.20 a^2$, $c=-1.09 a^3$.  In particular, we
note here that the sign of $c$ is the opposite of what we had previously
constrained it to be through our electric field ansatz.
\par
We now present the generalization of the electric field ansatz
$\Ev(\rv)$ in Eq.\ (\ref{electric1}) used to calculate the the results
discussed above and mention a few of the important properties it possesses.  We
insist that the generalization maintain the correct far-field behavior to
quadrupole order so that we can identify the quadrupole strength $c$, as well
as the dipole strength
$p_z$.  We would also like the generalization to put fewer constraints
on the allowed values of $p_z$ and $c$.  In particular, since in
certain situations the sign of $c$ seems to be an important quantity
we should certainly not restrict $c$ to be of a particular sign in the
ansatz.  Rewriting Eq.\ (\ref{electric1}) as,
\begin{equation}
{\Ev (\rv ) \over E_0} = \left(1-{a^3 \over r^3} \right)\ev_z +
\lambda^2 a^2 {\rv \over r^3} + {{3 z a^3} \over
r^5} \rv
\end{equation}
motivates the following generalization containing the aforementioned
desired properties,
\begin{equation}
{\Ev (\rv ) \over E_0} = \left( 1 - {a^{k_1} \over r^{k_1}} \right) \ev_z +
\lambda^2 a^2 {\rv \over r^3} + \left( { {\beta_1 a^3} \over r^5} + {
{\beta_2 a^{k_2}} \over r^{k_2+2}} \right) z \rv.
\end{equation}
As before $p_z=\lambda^2 a^2$ and now $c=\beta_1 a^3/2$,
provided that in carrying out the minimization over the appropriate
free parameters we find $k_1$ and $k_2$ are both greater than $3$ (if
this had not been the case then the far-field behavior would not have
been correct).  Minimizing we find an energy considerably lower than
that found using Eq.\ (\ref{electric1}). We obtain the results for
$U$, $z_0$, $p_z$ and $C$ given above.  In addition, we find
$k_1 = 4.88$, $k_2 = 3.73$, $\beta_2 = 4.27$.
\subsection{A Second Dipole Ansatz}
\par
To study the transition from a dipole to a Saturn ring and to
establish that the dipole has a lower energy than the Saturn ring, we
need an ansatz that allows disclination rings and a limiting
hyperbolic hedgehog.  To construct our ansatz we use appropriately
symmetric solutions to a related $2D$ problem and modify their
far-field behavior to match the required $3D$ far-field behavior (following
a route analogous to that in\cite{Terentjev95,KukRuh96} for the case of the
equatorial ring).
\par
To make contact with $2D$ configurations, it is convenient to look at
the general problem via the following parameterization,
\begin{eqnarray}
\nv & = & (\sin\Theta(\rv) \cos\Phi(\rv), \sin\Theta(\rv) \sin\Phi(\rv),
\cos\Theta(\rv))  \nonumber \\
\rv & = & (r \sin\theta \cos\phi,r \sin\theta \sin\phi,r \cos\theta)
\label{npolar}
\end{eqnarray}
where we have expressed $\rv$ in the usual spherical coordinates.
The full Euler-Lagrange equations in $\Theta,\Phi$ arising from the
free energy of Eq.\ (\ref{frank2}) are simply,
\begin{eqnarray}
\nabla^2 \Theta - \sin\Theta \cos\Theta (\nabla \Phi \cdot \nabla
\Phi) = 0 \nonumber \\
\sin\Theta \nabla^2 \Phi + 2 \cos\Theta (\nabla \Theta \cdot
\nabla \Phi) = 0 \nonumber
\end{eqnarray}
The nonlinearity of the above coupled partial differential equations (PDE's)
make closed-form solutions difficult to obtain, though we note here for
completeness that
$\Theta=\theta, \Phi=\phi$ is indeed a solution (the radial hedgehog)
and thus that solutions are not impossible to find.
\par
Turning now to the problem at hand, we should certainly impose the
condition of azimuthal symmetry, namely: $\partial_{\phi} \Theta=0 ,
\partial_{\phi} \Phi = 1$, and $\Theta |_{\theta=0,\pi}$ is either $0$
or $\pi$.  We will impose a more stringent constraint on $\Phi$, namely that
$\Phi = \phi$, which allows us to use our knowledge of $2D$ nematic
configurations to construct $3D$ azimuthally symmetric configurations
(this more stringent condition was also satisfied in the electric
field ansatz).  This leaves us with the single Euler-Lagrange equation,
\begin{equation}
\nabla^{2} \Theta - {\sin2\Theta \over {2 r^2 \sin2\theta} } = 0,
\label{eulag}
\end{equation}
whose nonlinearity still makes solutions difficult to obtain. Using
now that we want $\nv \rightarrow (0,0,1)$ (or $\Theta \rightarrow 0$) as
$r \rightarrow \infty$, we can linearize (\ref{eulag}) and find the form
of the far-field solutions,
\begin{eqnarray}
\Theta & \rightarrow & \sum_{k=1}^{\infty} \frac{A_k}{r^{k+1}}
P_k^1(\cos\theta) \nonumber \\
& & =  A_1 {\sin\theta \over r^2} + A_2 { {3 \sin2\theta} \over {2
r^3}} + \cdots \nonumber \\
& & \equiv  p_z {\sin\theta \over r^2 } + c {\sin2\theta \over
r^3} + \cdots,
\label{ThetaFF}
\end{eqnarray}
where the last line defines $p_z$ and $c$ to match the definitions of
these quantities in Eq. (\ref{multipole2}), as one can check by substituting
this form for $\Theta$ into Eq.(\ref{npolar}).
\par
Given the restriction, $\Phi=\phi$, we note that in the $x-z$ plane we
have,
\begin{equation}
\nv_{2 D} = (\sin\Theta, \cos\Theta).
\end{equation}
For $\nv_{2 D}$, we have a wealth of information about how to construct
solutions for the harmonic free energy,
\begin{equation}
F_{2 D} = \case{1}{2} K \int d^2 r (\nabla_{2 D} \Theta)^2.
\end{equation}
It would be nice if our problem reduced to this linear problem.  However,
even though this is not the case we can still exploit our knowledge of the $2
D$ solutions to construct ansatz solutions for the $3 D$ problem.  The
procedure is quite simple and has at least some promising motivations.
Any configuration of $\nv_{2D}$ that is invariant under $x\rightarrow -x$
can be converted to a $3D$ configuration by spinning about the $z$ axis to
produce
\begin{equation}
\nv = (\sin\Theta \cos\phi, \sin\Theta \sin\phi,\cos\Theta), \nonumber
\end{equation}
a solution with the $\Phi=\phi$ constraint.
\par
Furthermore we note that the canonical $q_{2D}=+1$ defect on the
$z-$axis becomes a $q=1$ radial hedgehog in $3D$, and the canonical
$q_{2D}=-1$ defect on the $z-$axis becomes a $q=1$ hyperbolic hedgehog
in $3D$ (recall that in $2D$ the charges of nematic defects are signed
and their composition law is addition).  Also, any symmetric pair (we need
a pair to maintain the required reflection symmetry about the
$z-$axis) of $q_{2D}=\pm\case{1}{2}$ defects off the
$z-$axis, when spun into a $3D$ configuration become a
strength $\case{1}{2}-$disclination ring.  Finally, we note that satisfying the
$3D$ BC on the sphere, namely that $\nv={\bf e_r}$ on the sphere, simply
requires satisfying normal BC on the circle in $2D$.
In $2D$ we can satisfy these BC using the method of images, which
works because of the linearity of the EL equations.
\par
Before writing down the ansatz we note that the $2D$ configuration
$\Theta=\theta_{z_0}$, where $\theta_{z_0}$ is the polar angle
measured with respect to the point $z_0$ on the $z-$axis, when made
into a $3D$ configuration indeed satisfies the $3D$ Euler-Lagrange (EL)
equations.  In two dimensions, $\Theta_1 + \Theta_2$ is a solution to the
EL equations provided both $\Theta_1$ and $\Theta_2$ are.  However, because
the $3D$ EL equations are nonlinear, $\Theta_1 + \Theta_2$ is in general
not a solution to the $3D$ El equations even if $\Theta_1$ and $\Theta_2$
individually are.  It is this fact that prevents our ansatz configurations
from being true solutions to the $3D$ equations.
\par
\begin{figure}[t]
\twocolumn[\hsize\textwidth\columnwidth\hsize\csname@twocolumnfalse\endcsname
\centerline{\epsfbox{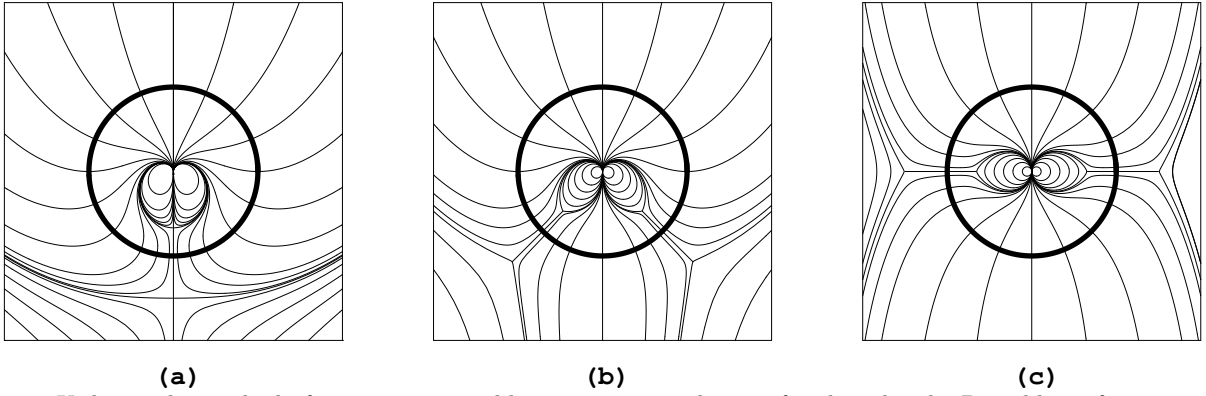}}
\caption{Utilizing the method of images we are able to construct
solutions for the related $2 D$ problem of a nematic with homeotropic
boundary conditions at infinity and with normal boundary conditions on
a circle about the origin.  These configurations are then extended to
$3 D$ configurations by spinning them about their vertical symmetry
axis, where the singularity along the vertical axis become
singularities in $3 D$ (possibly removable ones as in the case of the
$+2$ defect at the origin) and the symmetric pairs of defects off the
axis become singular defect rings.  Note that in our calculations we
are not really concerned with the from of the field inside the circle
since the nematic is only present in the exterior, the field is merely
drawn here to elucidate the origin of the ansatz used. (a) a
hyperbolic defect beneath the sphere.  (b) a non-equatorial
disclination ring.  (c) an equatorial disclination ring.}
\label{figthreedfcts}
\vskip 2pc]
\end{figure}

We now construct our ansatz for a sphere at the origin with a compensating
unit strength hyperbolic hedgehog.  To keep equations as simple as
possible, we use units in which the sphere radius is one.  As discussed
above, we first construct a solution to the two-dimensional problem.  This
is done using the fact that a defect of strength $q$ at position
$\rv_{2D} = (x_0, z_0 )$ is described by the
field $\Theta = q \tan^{-1} [(z-z_0)/ (x-x_0)]$.  Boundary conditions at
the sphere's surface and at infinity can be met by placing a strength
$q=+2$ defect at the sphere's center, a $-1$ defect at position $(0, - r_d )$
for arbitrary $r_d$, and an image $-1$ defect inside the spheres at
$(0, r_d^{-1})$ as shown in Fig.\ \ref{figthreedfcts}a.
This leads to
\begin{equation}
\Theta_0 = 2 \theta - \tan^{-1}{r\sin\theta \over {r\cos\theta +
r_d}} -\tan^{-1}{r r_d\sin \theta \over {r r_d\cos \theta + 1}}, \nonumber
\end{equation}
where we have taken the radius of the sphere to be $1$, and $r_d$ is the
distance from the defect below the sphere to the origin.  While this
does have the correct form near the defect at $z=-r_d$ it does not have
the correct far field form for the $3D$ problem.  In fact,
\begin{equation}
\Theta_0 \approx \left(r_d+{1\over r_d}\right) {\sin\theta \over r} -
\left(r_d^2 + {1\over r_d^2}\right) {\sin\theta\cos\theta \over r^2} + \cdots,
\label{Theta0FF}
\end{equation}
which does not agree with Eq.\ (\ref{ThetaFF}).  However, we note that to
the order shown, it differs only by an overall power of $r^{-1}$
(this is not true for the higher order terms not shown).  So, we alter
$\Theta_0$ in the following way, taking care to preserve the BC at
$r=1$.
\begin{eqnarray}
\Theta & & =  2 \theta - \tan^{-1}{r\sin\theta \over {r\cos\theta +
r_d}} - \tan^{-1}{r r_d\sin\theta \over {r r_d\cos\theta + 1}} \nonumber \\
& &  + e^{-k/r^3}\Biggl[\left(r_d+{1\over r_d}\right)
\left(-{1 \over r} + {1 \over r^2}\right) \sin\theta \nonumber \\
& &  +{1\over 2}\left(r_d^2+{1 \over r_d^2}\right)\left({1 \over r^2}-{1
\over r^3}\right) \sin2\theta \nonumber \\
& &  + {1\over 3} \left(r_d^3+{1 \over r_d^3}\right)\left(-{1 \over
r^3}+{1 \over r^4}\right) \sin\theta (4\cos^2\theta-1) \Biggr],
\label{Thetapt}
\end{eqnarray}
which now has the far-field form,
\begin{eqnarray}
\Theta \approx  \left(r_d + {1 \over r_d}\right){{\sin\theta} \over {r^2}} -
{1 \over 2}\left({1 \over r_d^2} + r_d^2\right){{\sin2\theta} \over
{r^3}} .
\label{ThetaptFF}
\end{eqnarray}
This agrees with Eq.\ (\ref{ThetaFF}) with $p_z = r_d + r_d^{-1}$ and
$c = -(r_d^2 + r_d^{-2} )/2$, which contrary to the electric field
ansatz is negative (recall $a=1$).  The factor $e^{-k/r^3}$, introduced
for numerical convenience,
tends to $1$ at large $r$ and to a value near the sphere controlled by the
variational parameter $k$.
Substituting this form [Eq.\ (\ref{Thetapt})] for $\Theta$ (and
$\Phi=\phi$) in Eq.\ (\ref{frank2})  and minimizing over $r_d$ and $k$
(numerically) we find, $k=0.32$ and $r_d=1.22$.
\par
Now taking the sphere to have a radius of $a$, as in the
electric-field ansatz, we find $p_z=2.04a^2$, $c=-1.08a^3$ and
$U=7.84 U_0$ in very good agreement with the results obtained from the
generalized electric field ansatz.  In addition $r_d =1.22a$ agrees nicely
with the corresponding quantity $z_0= 1.26a$ obtained from the electric
electric-field ansatz. Preliminary results obtained via numerical solution
to the full Euler-Lagrange equations for this problem are in excellent
agreement with these ansatz results\cite{StaSte97}.
\subsection{From Dipole to Saturn Ring}
\par
We can easily generalize the angular parametrization of a point defect just
discussed to describe an annular ring defect with a varying opening angle
$\theta_d$ [Fig.\ \ref{figthreedfcts}b] and thereby study the
transition from a dipole
configuration wth $\theta_d = \pi$ to the saturn ring with $\theta_d =
\pi/2$ [Fig.\ \ref{figthreedfcts}c].  We proceed exactly as in the
point-defect case.  We place one strength $+2$ defect at the center of the
circle, two strength $-1/2$ defects at $\rv_{2D} =
r_d ( \pm \sin \theta_d, \cos \theta_d )$, and two strength $-1/2$ images
inside the sphere at $\rv_{2D} = r_d^{-1} ( \pm \sin \theta_d , \cos
\theta_d )$.  This gives a solution to the $2D$ problem with two strength
$-1/2$ defects outside the circle.  When promoted to $3D$, this solution
correctly satisfies homeotropic boundary conditions at the surface of the
sphere, and it yields a strength $1/2$ disclination ring outside the sphere
with an opening angle of $\theta_d$.  It fails, however to produce the correct
far-field form for
$\nv$.  We add terms similar to those of the preceding calculation to
correct this deficiency to produce
\begin{eqnarray}
\Theta && =  2 \theta - {1 \over 2} \Biggl[ \tan^{-1}{{r\sin\theta - r_d
\sin\theta_d} \over {r\cos\theta - r_d \cos\theta_d}}  \nonumber \\
&& + \tan^{-1}{{r\sin\theta + r_d
\sin\theta_d} \over {r\cos\theta - r_d \cos\theta_d}}
 \nonumber \\
&& + \tan^{-1}{{r r_d\sin\theta -
\sin\theta_d} \over {r r_d\cos\theta -  \cos\theta_d}} +
\tan^{-1}{{r r_d\sin\theta +
\sin\theta_d} \over {r r_d\cos\theta -  \cos\theta_d}}\Biggr]
\nonumber \\
&& + e^{-k/r^3} \Biggl[\left(r_d + {1 \over r_d}\right)
\left({1 \over r} - {1\over r^2}\right)\cos\theta_d\sin\theta
\nonumber \\
&& + \left(r_d^2 + {1\over r_d^2}\right)
\left({1 \over r^2} - {1\over r^3}\right)(-1+2\cos^2\theta_d)
\sin\theta\cos\theta
\nonumber \\
&& + {1 \over 3}\left(r_d^3 + {1\over r_d^3}\right) \left({1\over r^3} -
{1\over r^4}\right)\cos\theta_d (4\cos^2\theta_d -3)\times \nonumber \\
&& \qquad \sin\theta(4\cos^2\theta-1)\Biggr] .
\label{ring}
\end{eqnarray}
The director configuration in the vicinity of the disclination ring is
singular, and the above form breaks down at distances from the disclination
ring less than the core radius $r_c$.
The equatorial ring configuration,
$\theta_d=\pi/2$ has been investigated
previously\cite{Terentjev95,KukRuh96}.
\par
\begin{figure}
\centerline{\epsfbox{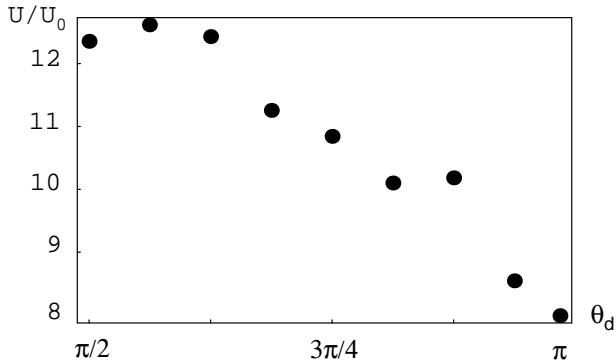}}
\caption{Energy versus angular position of the defect ring.  Note that
the equatorial ring ($\theta_d={\pi \over 2}$) does appear to enjoy
some metastability but that the collapsed ring (or effectively the
point defect, $\theta=\pi$) is of much lower energy.}
\label{figEvsthetad}
\end{figure}
\begin{figure}
\centerline{\epsfbox{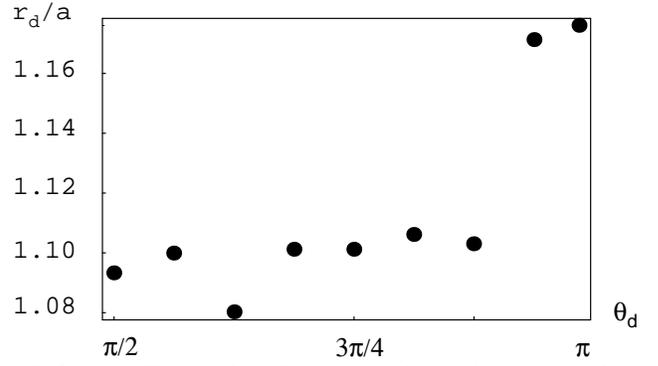}}
\caption{The preferred distance from the origin of the disclination
ring as a function of its angular position.}
\label{figrvsthetad}
\end{figure}

Figure \ref{figEvsthetad} shows the energy $U$ in units of $U_0 = \pi K
a/2$ for various values of $\theta_d$ obtained by minimizing the free
energy [Eq.\ (\ref{frank2})], augmented by an additional core energy $\pi
r_d K \sin \theta_d /2$, over the variational parameters $r_d$ and $k$
in the ansatz function Eq.\ (\ref{ring}). The reduced core radius in these
calculations was chosen to be $r_c = 10^{-3}$.  We have checked that our
results are insensitive to the value of $r_c$ for $10^{-4} < r_c < 10^{-2}$.
For a typical core radius of $10$nm, or results are thus good for
particles with radii as small as $1\mu$m.
Figure \ref{figrvsthetad} shows the corresponding equilibrium distance of the
ring from the center of the sphere.  We note that $r_d$ and $E$ for
$\theta_d \approx \pi$ agree quite well with the values obtained from
the point defect ansatz previously presented.  That is, this ansatz
does collapse down to the point defect in a nice manner.  However, we
must also note that for $\theta_d= \frac{\pi}{2}$, the equatorial
ring, our $r_d=1.08a$ is somewhat different from the value of $1.25a$
found by Terentjev in \onlinecite{Terentjev95,KukRuh96}.
\par
Further comments about the ring configuration are in order.
Recalling that in three dimensions the point defect
singularities represent integrable singularities whereas the ring
defects do not, one might naively expect that rings should always
collapse into points.  As discussed in Sec.\ \ref{Topdefects}, however,
this is not
always the case. Even an isolated point singularity might have
higher energy than an isolated ring
singularity\cite{MorNak88,Terentjev95,LavTer97} .  But it is
worth noting that in such circumstances the equilibrium radius of such
rings turns out to be quite small ($\approx 0.2\mu$m). Thus
here we would naively expect that if a ring configuration were to exist it
would probably not be an equatorial ring configuration.  This proves
to be correct as we see in Fig.\ \ref{figEvsthetad}. Though the equatorial ring
does appear to enjoy some metastability, its energy is considerably higher
than that of the point defect below the sphere.
\subsection{Thermal Stability}
\par
We have seen so far that the point defect beneath the sphere is the
energetically favorable configuration.  The elastic constant $k$ for
deviations $\Delta z$ from equilibrium separation $z_0$ is simply the
curvature of the energy versus separation curve
[e.g., Fig.\ \ref{figEfield1}]. The dipole ansatz yields $k = 33 \pi K/a$.
The other ansatzes yield similar values.  Thus
\begin{equation}
\left\langle\left({\Delta z \over z_0}\right)^2\right\rangle
 = {k_B T \over k z_0^2} \approx 10^{-5} ,
\end{equation}
where the final numerical estimate follows from $k_B T \approx
10^{-13}$erg, $K\approx 10^{-6}$dyne, and $a=1 \mu$m.
These fluctuations in the length
of the topological dipole are unobservably small.
\par
We have argued that the topological dipole prefers to align parallel or
anti-parallel to the director at infinity.  We will now show that the
angular restoring force constant $k_{\theta}$ is greater that $2 \pi K a$
so that
\begin{equation}
\langle (\Delta \theta )^2 \rangle  < {k_B T \over 4 \pi K a } \approx
10^{-4} .
\end{equation}
Thus, though of order ten times greater than fluctuations in the length of
the dipole, angular fluctuations are still unobservably small.
Interestingly, we note that angular fluctuations in the $2D$ version of
this problem are much larger and have indeed been observed in free
standing smectic films\cite{LinCla97}
\par
Our approach is to provide an ansatz for a director configuration with
the dipole rotated through an angle $\Delta \theta$ relative to the
director field at infinity given a configuration in which the dipole
moment is parallel to the director field at infinity. We will then use
this ansatz to calculate bounds on $k_{\theta}$.  We start with an
aligned dipole configuration with $\nv$ expressed in polar coordinates
according to Eq.\ (\ref{npolar}).
We then construct a rotated configuration, $\nv^{\prime}$, by slowly
rotating $\nv$ about the $y$-axis as we progress radially outward:
\begin{eqnarray}
\nv_{x}^{\prime} &=& \sin\Theta\cos\Phi\cos f + \cos\Theta\sin f
\nonumber \\
\nv_{y}^{\prime} &=& \sin\Theta\sin\Phi \nonumber \\
\nv_{z}^{\prime} &=& \cos\Theta\cos f - \sin\Theta\cos\Phi\sin f
\end{eqnarray}
where the amount we rotate at each point is given by the function
$f(\rv)$ which must satisfy the boundary conditions,
\begin{eqnarray}
f(r = a) &=& 0 \nonumber \\
f(r = R) &=& \Delta\theta.  \label{bc3}
\end{eqnarray}
$\Delta\theta$ denotes the tilt angle of the dipole with respect to
the far-field, $a$ the droplet radius, and $R$ the system size.  To
see that $\Delta\theta$ is the stated tilt angle we note that the
far-field of $\nv^{\prime}$ makes an angle $\Delta\theta$ with the
$z-$axis, and furthermore, the transformation of $\nv$ to
$\nv^{\prime}$ does not change the position of the singularity, or of
the droplet itself.  Thus, the droplet-defect dipole (${\bf p}$) is
still aligned with the $z-$axis.
\par
We denote $F(\Delta\theta)$ as the free energy of the tilted dipole
configuration and, accordingly refer to $F(0)$ as the free energy of
the aligned configuration. Using Eq.(\ref{frank2}) we find,
\begin{eqnarray}
& & F(\Delta\theta)-F(0) = \nonumber \\
& \case{1}{2} K& \int d^3 r
(\cos^2\Phi\sin^2\Theta + \cos^2\Theta)\gradv f \cdot \gradv f
\end{eqnarray}
where we have eliminated terms linear in $f$ using the symmetry
$F(\Delta\theta) = F(-\Delta\theta)$. Now noting that the
parenthetical term is always less than unity we have,
\begin{equation}
F(\Delta\theta)-F(0) > \case{1}{2} K \int d^3 r \gradv f \cdot \gradv
f. \label{esf1}
\end{equation}
The right hand side of this equation is a minimum when $\nabla^2 f = 0$
subject to the boundary conditions of Eq.\ (\ref{bc3}), i.e., when
\begin{equation}
f = \Delta \theta {1 - ( a/r) \over 1 - (a/R)} .
\end{equation}
Using this $f$ in Eq.\ (\ref{esf1}), we obtain
\begin{equation}
F(\Delta\theta) - F(0) > 2 \pi K (\Delta\theta)^2 a ,
\end{equation}
for $R \gg a$, implying $k_\theta > 4 \pi Ka$.
\subsection{Optical Images\label{optimages}}
In the experiments reported by Poulin {\em et al.\/} \cite{PouSta97}, only the
dipole and not the saturn ring configuration shown in
Fig. \ref{figdroplet1} is observed.
Figure \ref{dipimage}a presents an experimentally obtained image of a
single water droplet under crossed polarizers with one polarizer parallel
to the dipole axis. In the region of the
droplet we see a pronounced pattern arising from the spatially varying director
field. In Fig. \ref{dipimage}b we show an image of a similar single
droplet calculated using the Jones matrix
formalism\cite{Drzaic95} and neglecting any refraction at the droplet
boundary. The similarity of the two images is obvious and clearly confirms
the occurrence of the dipole configuration. Both pictures show two bright
wings left of the droplet. In the calculated picture they are much more
extended than they are in the experimental picture.
The theoretical picture was calculated using the simple electric field
ansatz.  The more sophisticated ansatzes reduce the region around the
defect where there is rapid variation of the director.  They would yield
images in closer agreement with the experimentally observed one.
\begin{figure}
\caption{(a) Image of a single droplet with its companion defect as
observed under crossed polarizers obtained by P. Poulin.  (b) Simulated
image of the same configuration using the Jones matrix formalism.  The two
images are very similar.}
\label{dipimage}
\end{figure}
\section{Phenomenological Theory and Droplet-Droplet Interactions}
\par
To understand the properties of multi-droplet emulsions, we need to determine
the nature of droplet-droplet interactions.  These interactions are mediated
by the nematic in which they are embedded and are in general quite
complicated.  Because interactions are determined by distortions of the
director field, there are multi-body as well as two-body interactions.
We will content ourselves with calculations of some properties
of the effective two-body interaction.  To calculate the
position-dependent interaction potential between two droplets, we should solve
the Euler-Lagrange equations, as a function of droplet separation, subject to
the boundary condition that the director be normal to each droplet.  Solving
completely these nonlinear equations in the presence of two droplets is
even more complicated than solving them with one droplet, and again we must
resort to approximations.  Fortunately, interactions at large separations
are determined entirely by the far-field distortions and the multipole
moments of the individual droplet-defect pairs, and they can be
described by a phenomenological free energy, which we will derive in this
section.
\par
In the preceding section, we established that each water droplet creates a
hyperbolic hedgehog to which it binds tightly to create a stable topological
dipole.  The original droplet is described by three translational degrees of
freedom.  It draws out of the nematic a hedgehog, which itself has three
translational degrees of freedom.  The two combine to produce a dipole with
six degrees of freedom, which can be parametrized by three variables
specifying the position of the water droplet, two angles specifying the
orientation of the dipole, and one variable specifying the magnitude of the
dipole.  As we have seen, the magnitude of the dipole does not fluctuate much
and can be regarded as a constant.  The direction of the dipole is also
fairly strongly constrained.  It can, however, deviate from the direction
of local preferred orientation (parallel to a local director to be defined in
more detail below) when there are many droplets present.  The droplet-defect
pair is in addition characterized by its higher multipole moments.  The
direction of the principal axes of these moments is specified by the direction
of the dipole as long as director configurations remain uniaxial.  The
magnitudes of the uniaxial moments like the magnitude of the dipole moment are
energetically fixed.  When director configurations are not uniaxial,
multipole tensors will develop additional components, which we will not
consider here.  We can thus parametrize droplet dipoles by their position and
orientation and a set of multipole moments, which we regard as fixed.
Let $\ev^{\alpha}$ be the unit vector specifying the direction of the dipole
moment associated with droplet $\alpha$.
Its dipole and quadrupole moments are then
$\pv^{\alpha} = p \ev^{\alpha}$ and $c_{ij}^{\alpha} = c ( e^{\alpha}_i
e^{\alpha}_j -\case{1}{3} \delta_{ij})$, where $p$ and $c$ are the magnitudes
of the dipole and quadrupole moments calculated in the preceding section.  We
can now introduce dipole- and quadrupole-moment densities, $\Pv( \rv)$ and
$C_{ij} (
\rv )$ in the usual way.  Let $\rv^{\alpha}$ denote the position of droplet
$\alpha$, then
\begin{eqnarray}
\Pv ( \rv ) & = & \sum_{\alpha} \pv^{\alpha}\delta ( \rv - \rv^{\alpha} )
\nonumber \\
C_{ij}(\rv) & = & \sum_{\alpha} c_{ij}^{\alpha} \delta ( \rv - \rv^{\alpha} ) .
\end{eqnarray}
\par
We now construct an effective free energy for director and droplets valid at
length scales large compared to droplet dimensions.  At these length scales,
we can regard the droplets as point objects (as implied by the definitions of
the densities given above).  At each point in space, there is a local
director $\nv ( \rv)$ along which the droplet-dipoles wish to align.  In the
more microscopic picture, of course, the direction of this local director
corresponds to the far-field director $\nv_0$.  The effective free energy is
constructed from rotationally invariant combinations of $P_i$, $C_{ij}$, $n_i$,
and the gradient operator $\nabla_i$ that are also even under $\nv
\rightarrow - \nv$.  It can be expressed as a sum of terms
\begin{equation}
F = F_{\nv} + F_{p} + F_{C} + F_{\rm align} ,
\end{equation}
where $F_{\nv}$ is the Frank free energy, $F_p$ describes
interactions between $\Pv$ and $\nv$, $F_{C}$ describes interactions
between $C_{ij}$ and $\nv$ involving gradient operators, and
\begin{eqnarray}
F_{\rm align}& =& - D \int d^3 r C_{ij} (\rv)  n_i (\rv) n_j ( \rv )
\nonumber \\
&= & - DQ \sum_{\alpha} \left[(\ev^{\alpha} \cdot \nv ( \rv^{\alpha}) )^2 -
\case{1}{3} \right]
\end{eqnarray}
describes the alignment of the axes $\ev^{\alpha}$ along the local director
$\nv ( \rv^{\alpha} )$. The leading contribution to $F_{p}$ is
identical to that for electric dipoles in a nematic\cite{Meyer69,DeGPro93}
\begin{equation}
F_p =4 \pi  K \int d^3 [ - \Pv \cdot \nv (\gradv \cdot \nv ) + \beta \Pv \cdot
(\nv \times \gradv \times \nv) ] ,
\label{Fp}
\end{equation}
where $\beta$ is a material-dependent unitless parameter.
The leading contribution to $F_C$ is
\begin{eqnarray}
F_C& =&4 \pi  K \int d^3 r [ (\gradv \cdot \nv) \nv \cdot \gradv (n_i C_{ij}
n_j)
\nonumber\\
& & \qquad + \gradv (n_i C_{ij} n_j) \cdot (\nv \times \gradv\times\nv )] .
\label{F_C}
\end{eqnarray}
There should also be terms in $F_C$ like $C_{ij}\nabla_k n_i \nabla_k
n_j$.  These terms can be shown to make contributions to the effective
droplet-droplet interaction that are higher order in separation than those
arising from Eq.\ (\ref{F_C}).  Equation (\ref{F_C}) is identical to that
introduced in Ref.\ \onlinecite{RamRaj96} to discuss interactions between
saturn-ring droplets, provided $n_i C_{ij} n_j$ is replaced by a scalar
density $\rho (\rv ) = \sum_{\alpha} \delta ( \rv - \rv^{\alpha} )$. The
two energies are absolutely equivalent to leading order in the components
$n_{\mu}$ of $\nv$ perpendicular to $\nv_0$ provided all $\ev^{\alpha}$ are
restricted to be parallel to $\nv_0$.  When this restriction on
$\ev^{\alpha}$ is lifted, and to higher order in $n_{\mu}$, the two
theories differ.  In our opinion, the scalar variable cannot strictly speaking
be used, because each droplet carries with it an anisotropic director
environment, even when the dipole moment is zero.
\par
\begin{figure}
\centerline{\epsfbox{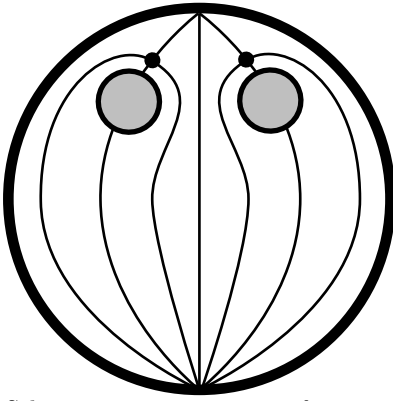}}
\caption{Schematic representation of two water droplets
with homeotropic boundary
conditions at their outer surfaces in a nematic drop with tangential
boundary conditions at its outer surface.  The total hedgehog charge
is zero, so there is one hyperbolic hedgehog per droplet.  The water
droplets migrate to the region of maximum splay near a surface boojum.  the
splay near the boojum is assumed to be sufficiently strong
that both dipoles prefer
to be near the boojum rather than to form a chain.}
\label{figtang}.
\end{figure}

Since $\Pv$ prefers to align along the local director $\nv$, the
dipole-bend coupling term in Eq.\ \ref{Fp} can be neglected to leading
order in deviations of the director from uniformity.  The $- \Pv \cdot \nv
(\gradv \cdot \nv )$ term in Eq.\ (\ref{Fp}) shows that dipoles aligned along
$\nv$ create local splay as is evident from the dipole configuration
depicted in Fig.\ \ref{figdroplet1}a.  In addition, this term says that
dipoles can lower their energy by migrating to regions of maximum splay
while remaining aligned with the local director.
Experiments\cite{PouSta97,PouWei97} support this conclusion.  Water
droplets in a nematic drop with homeotropic boundary conditions at its
surface congregate at the center of the nematic
drop where the splay is a maximum.
Boundary conditions at the outer surface of nematic drops
can be changed from homeotropic to
tangential by adding a small amount of glycerol to the continuous water
phase.  In the passage from homeotropic to tangential boundary conditions,
the topological charge of the nematic drop changes from one to zero, and
point defects called boojums\cite{Mermin77,CanLeR73,KurLav82} form on the drops
surface.  The director splay is a maximum in the drop's interior near the
boojums.  Water droplets move from the drop centers to drop surfaces near
boojums as the boundary conditions are changed.  The final configuration of
two droplets in a nematic drop with tangential boundary conditions is shown
in Fig.\ \ref{figtang}.
\par
To harmonic order in $n_{\mu}$, the full effective free energy is
\begin{equation}
F = K\int d^3 r [\case{1}{2}(\gradv n_{\mu})^2 - 4 \pi P_z \partial_{\mu}
n_{\mu} + 4 \pi (\partial_z C_{zz} ) \partial_{\mu} n_{\mu}] .
\label{fpchar}
\end{equation}
The dipole-bend coupling term of Eq.\ (\ref{Fp}) does not contribute
because $\Pv$ is aligned along the far-field director.
Thus,
\begin{equation}
\nabla^2 n_{\mu} = 4 \pi \partial_{\mu}[ P_z (\rv) -\partial_z
C_{zz}(\rv) ],
\label{npchar}
\end{equation}
or
\begin{equation}
n_{\mu} ( \rv ) = - \int d^3 r^{\prime} {1 \over |\rv - \rv^{\prime}|}
\partial_{\mu}^{\prime} [ P_z (\rv^{\prime}) - \partial_z^{\prime} C_{zz}(
\rv^{\prime}) ] .
\label{npchar2}
\end{equation}
For a single droplet at the origin with $\ev =\nv_0$, $P_z (
\rv) = p_z \delta (\rv)$ ($p_z = \pm p$), and $C_{zz}(\rv ) = \case{2}{3} c
\delta (\rv)$, and the above equation yields exactly Eq.\
(\ref{multipole2}).
\par
\begin{figure}
\centerline{\epsfbox{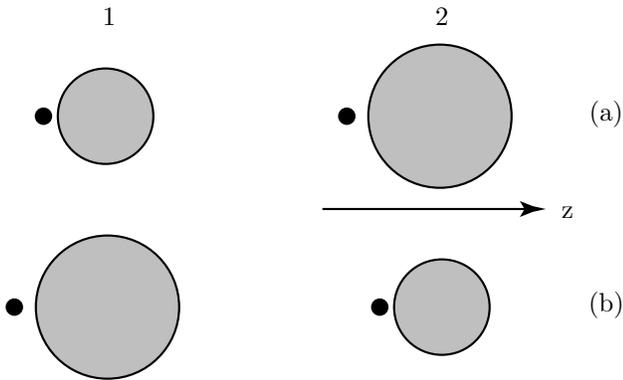}}
\caption{(a) configuration in which the large particle is to the right of
the small particle. (b) the inverse configuration.  The force between the
two particles is more attractive in case (a) than in case (b).  In both
cases, the particle to the left is labeled $1$ and that to the right $2$}
\label{figlargesm}
\end{figure}

Droplets create far-field distortions of the director, which to leading order
at large distances are determined by Eq.\ (\ref{npchar}), that interact with
the director fields of other droplets.  This leads to an effective
droplet-droplet interaction that can be expressed to leading order as
pair-wise interactions between dipole and quadrupole densities.
Using Eq.\ (\ref{npchar2}) in Eq.\ (\ref{fpchar}), we obtain
\begin{eqnarray}
{F\over 4 \pi K} &=&{1 \over 2} \int d^3 r d^3 r^{\prime}[ P_z ( \rv )
V_{PP} (\rv -\rv^{\prime}) P_z ( \rv^{\prime} ) \nonumber \\
& & \qquad \qquad+ C_{zz} ( \rv )V_{CC} (\rv-\rv^{\prime} ) C_{zz}
(\rv^{\prime} )
 \\
& & \qquad + V_{PC} ( \rv - \rv^{\prime} ) [
C_{zz} ( \rv ) P_{z} (
\rv^{\prime}) - P_z(\rv ) C_{zz} ( \rv^{\prime}) ] \nonumber
\end{eqnarray}
where
\begin{eqnarray}
V_{PP}(\rv) & = &\partial_{\mu} \partial_{\mu} {1 \over r} = {1\over r^3} ( 1
- 3 \cos^3 \theta ) \nonumber \\
V_{CC}(\rv) & = & - \partial_z^2 \partial_{\mu}\partial_{\mu} {1 \over r} =
{1 \over r^5}( 9 - 90 \cos^2 \theta + 105 \cos^4 \theta ) \nonumber \\
V_{PC}(\rv) & = & \partial_z \partial_{\mu} \partial_{\mu} {1 \over r} = {\cos
\theta \over r^4} ( 15 \cos^2 \theta - 9 ) ,
\end{eqnarray}
where $\theta$ is the angle the separation vector $\rv$ makes with $\nv_0$.
The interaction energy between droplets at positions $\rv$ and
$\rv^{\prime}$ with respective dipole and quadrupole moments $p_z$,
$p_z^{\prime}$, $c$ and $c^{\prime}$ is thus
\begin{eqnarray}
U( \Rv )& = &4 \pi K \left[ p_z p_z^{\prime} V_{PP} ( \Rv ) +
{4 \over 9}c c^{\prime} V_{CC}
( \Rv ) \right. \nonumber \\
& & \left. \qquad {2 \over 3}(c p_z^{\prime} - c^{\prime} p_z ) V_{PC}( \Rv )
\right] .
\end{eqnarray}
This potential can be used to calculate the force between two droplets
as a function of their separation.  Consider, for example, the interaction
between two droplets labeled $1$ and $2$ with respective radii $a_1$ and
$a_2$.  For simplicity, assume the dipoles associated with each droplet are
aligned along the positive $z$ axis and that the center of droplet $1$ is
at the origin and that of droplet $2$ at $\rv = (0,0,R)$ a distance $R$
away along the positive $z$ axis as shown in Fig.\ \ref{figlargesm}.
The dipole and quadrupole moments scale
respectively as $a^2$ and $a^3$, and we can write $p_z = \alpha a^2$ and
$c = - 3 \beta a^3/2$.  The dipole ansatz solution of Sec. IVC, predicts
$\alpha = 2.04$, nd $\beta = (2/3)\times 1.08 = 0.72$. The force between
two droplets is then
\begin{eqnarray}
{F\over 4 \pi K}& =& -  \alpha^2 a_1^2 a_2^2 {6 \over R^4}
+ \beta^2 a_1^3 a_2^3 {120\over R^6} \nonumber \\
& & - \alpha  \beta a_1^2 a_2^2 (a_1 - a_2 ){24 \over R^5} .
\end{eqnarray}
The dominant force is the attractive dipole-dipole force proportional to
$R^{-4}$.  Recent experiments confirm this relation\cite{PouWei97b}.
Interestingly the sign of the dipole-quadrupole force, which dies off as
$R^{-5}$, vanishes for particles of equal radius.  When the particle have
unequal radii, the sign of this force depends on the relative position of
the large and small particle.  If $a_1< a_2$, it is repulsive (for
$\beta>0$); if $a_1>a_2$, it is attractive, i.e., it is repulsive if the
smaller ball is to the right (positive $z$) of the large ball and
attractive if it is to the left (negative $z$).
\section{Summary and Conclusions}
\par
Inverse nematic emulsions in which surfactant coated water droplets are
dispersed in a nematic host have properties that are distinct from those
found in colloids, emulsions of two isotropic fluids, and of emulsions of
nematic droplets in an isotropic fluid.  The water droplets in these
emulsions exhibit anisotropic interactions that are repulsive at short range
and attractive at long range.  The short-range repulsive interaction
prevents coalescence of droplets and leads to long-term stability, which
can be eliminated by heating into the isotropic phase.  The long-range
attractive force is dipolar and favors chaining of droplets.
\par
In this paper, we have presented a detailed theoretical study of droplets
and droplet interactions in inverse nematic emulsions.  Homeotropic
boundary conditions at droplet surfaces produce a hedgehog director
configuration around each droplet.  Constraints on the
global topological charge force the nucleation of compensating topological
defects out of the nematic host.  The compensating defect associated with
a single droplet in a cell with a parallel aligned director at infinity can
be a point hedgehog or a disclination ring sitting above or below the
droplet or encircling its equator in the saturn ring configuration as shown
in Fig.\ \ref{figdroplet1}.  Using various variational ansatzes, we showed
that in the lowest energy configuration, a single water droplet pulls a
single point hedgehog from the nematic to form a tightly bound dipole.
Then, using a phenomenological model in which the topological dipoles
are coupled to the nematic director via a flexoelectric interaction, we
derived the effective long-range dipolar interaction between water
droplets.  We also considered quadrupolar corrections to the dominant
dipolar interaction. The phenomenological model also predicts the
experimentally observed tendency of dipoles to seek regions of high splay.
\par
We have focussed mostly on interactions between droplets in cells with
parallel boundary conditions at infinity with total topological charge
zero.  Multiple emulsions in which water droplets are dispersed in nematic
drops, which are in turn dispersed in water, have made possible the
isolation of a finite number of droplets and facilitated a number of
experimental observations.  The nematic drops are characterized by a
topological charge of one rather than zero and by spatially nonuniform
director configurations.  Many of the properties of these
droplets-within-drops systems such as chaining and the tendency of the
water droplets to concentrate near the center of the nematic drop are
explained by the analyses in this paper.  Numerically accurate predictions
about these systems, however, require, global minimization procedures that
can only be done numerically.  Numerical algorithms to study droplets
dispersed in confined geometries are currently under
development\cite{StaSte97}.
\par
Inverse nematic emulsions are a relatively new addition to the ever growing
list of interesting soft materials, and they offer the hope of new and
surprising properties.  We are currently investigating among other things
the dynamics of droplets in inverse emulsions and inverse emulsions of
water droplets in cholesteric rather than nematic liquid crystals.

\acknowledgments
We are grateful for the close working relation with Philippe Poulin and
David Weitz, who carried out the experiments described in this paper and
without whom this work would not have been done.
We are particularly grateful to Phillipe Poulin for providing with Fig.
\ref{dipimage}a and to Robert Meyer for helpful discussions and for
suggesting the director donfiguration shown in Fig.\ \ref{figdroplet2}b.
We also
acknowledge helpful discussions with Arjun Yodh and Martin Zapotocky.
T.C.L. and D. P. were supported primarily by the Materials Research Science and
Engineering Center Program of NSF under award number DMR96-32598.
H.S. Acknowledges a grant from the Deutsche Forschungsgemeinschaft under
grant number Sta 352/2-2.
\bibliography{/u/tom/datab/physics,/u/tom/datab/tgb,/u/tom/datab/liqcryst}
\bibliographystyle{prsty}

\end{document}